\documentclass[reprint,bibnotes,longbibliography,amsmath,amssymb,aps,prl,preprintnumbers]{revtex4-2}
\usepackage[T1]{fontenc}
\usepackage{bm,booktabs,microtype,amsthm}
\usepackage{dsfont}
\usepackage[colorlinks=true,linkcolor=blue,citecolor=blue,urlcolor=blue]{hyperref}
\newcommand{\Sh}{\operatorname{Sh}}
\newcommand{\Stir}[2]{\genfrac{[}{]}{0pt}{}{#1}{#2}}
\newcommand{\cA}{\mathcal A}
\newcommand{\Sym}{\operatorname{Sym}}

\newcommand{\e}{\mathrm e}

\begin{document}
\preprint{MPP-2026-200}
\title{Collinear Gluons from Gravity:\\[1mm] Poincar\'e-Birkhoff-Witt Decomposition and Twisted Intersections}

% Subleading Collinear Gluon Amplitudes from Gravity:\\[1mm]
% Poincar\'e-Birkhoff-Witt Reconstruction and Twisted  Intersections
\author{Jin Dong}
\email{jindong@mpp.mpg.de}
\author{Stephan Stieberger}
\email{stephan.stieberger@mpp.mpg.de}
\affiliation{Max-Planck-Institut f\"ur Physik, Werner-Heisenberg-Institut, Boltzmannstr. 8, 85748 Garching bei M\"unchen, Germany}
\begin{abstract}
We give an explicit reconstruction of subleading collinear Yang--Mills amplitudes from one-graviton amplitudes in Einstein--Yang--Mills theory and its string corrections.  We realize the monodromy kernel in the mixed open-closed string relation as a weighted Lie-algebraic recursion on words of gluon labels. At order $\alpha'^r$, its output contains at most $r$ Lie factors. The Poincar\'e--Birkhoff--Witt (PBW) theorem decomposes the space spanned by gluon orderings into sectors whose dimensions are given by unsigned Stirling numbers of the first kind. Appropriate shuffle sums of one-graviton amplitudes extract the degree-$r$ sector, for which the string-period matrix drops out. Accordingly, we show that the reconstruction coefficients factor into linear kinematic factors over the product of bipartition factors. Finally, we express the reconstruction coefficients as twisted intersection numbers between Parke--Taylor forms and explicit forest cocycles constructed recursively from logarithmic four-point differentials. %Finally, we also derive KK-like relations among one-graviton amplitudes at each fixed order in $\alpha'$.
\end{abstract}
\maketitle

\textbf{Introduction.---}
String theory provides a worldsheet origin for relations
between gauge and gravitational scattering amplitudes. One prominent example are the Kawai--Lewellen--Tye (KLT) relations expressing gravity amplitudes on the worldsheet sphere \cite{Kawai:1985xq} and torus \cite{Stieberger:2022lss} in terms of squares of gauge amplitudes.
On the worldsheet disk, contour deformations express mixed
open--closed string amplitudes as linear combinations
of pure open-string amplitudes, with coefficients
determined by monodromy
phases~\cite{Stieberger:2009hq,Stieberger:2015vya}.
For a massless gravitational insertion, these relations
represent a graviton by two collinear gauge bosons
sharing its momentum~\cite{Stieberger:2014cea}.
Related constructions express tree-level graviton
amplitudes through collinear limits of Yang--Mills (YM)
amplitudes~\cite{Stieberger:2015qja}, while mixed-disk
monodromies also yield relations among
Einstein--Yang--Mills (EYM) amplitudes and their
expansions in pure YM
amplitudes~\cite{Stieberger:2016lng}.

These relations probe collinear behavior beyond leading factorization (see also~\cite{Nandan:2016ohb}). For two adjacent collinear gluons, the leading singular term of a color-ordered
YM amplitude is fixed by a universal splitting factor multiplying a lower-point amplitude. The finite subleading terms contain additional information: particular combinations are related to one-graviton
EYM amplitudes in which the graviton replaces the collinear pair~\cite{Stieberger:2015kia}.

The converse problem is more demanding. A one-graviton EYM  amplitude determines a combination of YM orderings, rather than each ordering separately. Ordinary EYM amplitudes therefore do not necessarily contain sufficient information to reconstruct the full subleading collinear data. In~\cite{Dong:2026lsb}, we proposed a resolution by introducing higher-derivative gravitational string corrections~\cite{Schlotterer:2016cxa} together with homogeneous Bern--Carrasco--Johansson~\cite{Bern:2008qj,Bern:2010ue} (BCJ)-like relations. The successive numbers of new equations were conjectured to be unsigned Stirling numbers of the first kind, as checked up to $n=9$. Complementary recursion methods construct the collinear YM coefficients directly~\cite{Dong:2026kys}. These results motivate constructing an explicit inverse for arbitrary multiplicities.

String corrections might appear to make this inversion increasingly complicated. Open-string disk amplitudes organize their dependence on the inverse string tension $\alpha'$ in a matrix of worldsheet integrals multiplying a YM basis~\cite{Mafra:2011nv,Mafra:2011nw}. However, the algebraic structure that simplifies the inversion is already visible in the monodromy factors. Their relative phases turn insertion of a gluon label into a weighted commutator in the free associative algebra of words. In this letter, we use the Poincar\'e--Birkhoff--Witt (PBW) theorem to organize these products by their number of Lie factors (c.f.~\cite{Frost:2020eoa} for the Lie bracket in momentum kernel). The monodromy recursion bounds that number at each order in $\alpha'$. At order $\alpha'^r$, contributions involving
nontrivial string-period coefficients have PBW degree below $r$ and are therefore removed by projection onto degree $r$. Appropriate shuffle sums extract this component directly from physical gravitational amplitudes, and the inverse coefficients factor into linear kinematic terms. Accordingly, we construct the solution at arbitrary multiplicity, turning the proposed gravitational description into a direct reconstruction formula.

The reconstruction also admits a geometric description
in terms of twisted intersections on ${\cal M}_{0,n}$.
We associate a logarithmic cocycle to each PBW block
decomposition by recursively combining four-point
differentials that treat the two collinear punctures
symmetrically.
Their intersections with Parke--Taylor forms reproduce
the reconstruction coefficients, relating the PBW
organization of gravitational sources to the geometry
of the punctured world--sheet.

\textbf{The mixed open-closed string relations and the counting conjecture.---}Let $n\geq5$ count particles before fusion, $m=n-3$, $Y=\{2,\ldots,n-2\}$, and $G=\{1\}\cup Y$.The momenta of the collinear pair are taken to be $p_{n-1}=p_n=P/2$, with $P^2=0$, and the two particles are assigned equal polarizations. With $S_Y$ denoting the permutations of $Y$, the $m!$ unknown  subleading collinear data are
\begin{equation}
 C_\rho=A_{\rm YM}(1,\rho,n-1,n)_{\rm sub},\quad\rho\in S_Y.
 \label{C}
\end{equation}
We adapt the symmetric strict collinear prescription of Refs.~\cite{Dong:2026lsb,Dong:2026kys}. In this prescription the additional finite contribution of the adjacent factorization channel vanishes. The one-graviton (string) amplitude $A(1,\sigma;h;\alpha')$ has gluons $G$ and graviton $h$ with momentum $p_{n-1}+p_n=P$. Its low-energy expansion is written as $A=\sum_{r\geq1}\alpha'^r A^{(r)}$ where the superscript denotes the absolute $\alpha'$ order. Thus $r=1$ corresponds to the ordinary EYM amplitude, while $r=3$ gives the first non-vanishing string correction. Throughout, we strip the overall couplings and keep powers of $\pi$ explicit.

We define the Mandelstam variables in the collinear limit, subject to momentum conservation as
\begin{equation}
 s_{ij}=2p_i\!\cdot p_j,\quad s_{iP}=-\sum_{j\in G\setminus\{i\}}s_{ij},
 \quad\sum_{i<j\in G}s_{ij}=0.
 \label{kin}
\end{equation}
The mixed open--closed string relation is the starting point of the construction. For the canonical gluon ordering, the mixed string
relation of Ref.~\cite{Stieberger:2015vya}
takes the following form in the symmetric collinear limit:
\begin{equation}
 \begin{aligned}
 &A(1,2,\ldots,n-2;h;\alpha')\\[-1mm]
 &\quad=\sum_{l=2}^{n-2}(-1)^{n-2-l}
 \sin\!\left(\frac{\pi\alpha'}2s_{lP}\right)\\[-1mm]
 &\qquad\times\sum_{\rho\in\Omega_l}e^{i\pi\alpha'\Phi_l(\rho)}
 A_{\rm open}^{\rm strict}(1,\rho,n-1,n).
 \end{aligned}
 \label{mixed-disk}
\end{equation}
Here $\Omega_l$ consists of shuffles of the words $(2,\ldots,l-1)$ and $(n-2,\ldots,l+1)$, followed by $l$. A shuffle preserves the order within each word; either word may be empty. We also define $a\prec_\rho b$ when $b$ appears after $a$ in the ordering $\rho$. The phase factor is
\begin{equation}
 \Phi_l(\rho)=\frac{s_{lP}}2+
 \sum_{a=l+1}^{n-2}\left(s_{aP}+
 \sum_{\substack{b\in Y\\a\prec_\rho b}}s_{ab}\right).
 \label{disk-phase}
\end{equation}
The combined sign, sine, and exponential in Eq.~\eqref{mixed-disk} constitute the monodromy factor: they encode the contour rearrangement that replaces the closed-string insertion by two open-string insertions. Moreover, the open-string amplitudes can be reduced to a Yang--Mills basis through the string period matrix $F(u)$~\cite{Mafra:2011nv,Mafra:2011nw}, namely, $\bm A_{\rm open}^{\rm strict}=F(u)\bm C$, where $u=i\pi\alpha'$ and $F(0)=\mathds{1}$. Note that the superscript ``strict'' indicates that the same collinear prescription is imposed on both sides at subleading order. In particular, the right-hand side contains only terms with finite kinematic prefactors multiplying the finite subleading YM amplitudes.

Permuting the labels in $Y$ gives $m!$ relations of the form \eqref{mixed-disk}. Collect their left-hand sides into $\bm A$, and denote the matrix of monodromy factors by $J(u)/i$. The string relations become
\begin{equation}
 \bm A(\alpha')=\frac1i J(u)F(u)\bm C,
 \qquad u=i\pi\alpha',\qquad F(0)=\mathds{1}.
 \label{input}
\end{equation}
The rows of $J(u)F(u)$ label gravitational orderings and columns label the adjacent YM basis. The period matrix acts from the right and remains regular in the strict collinear limit, as its $\alpha'$-expansion contains no massless poles. We define $\mathcal K_r=[\alpha'^r](JF/i)$ as the full complex coefficient matrix at order $\alpha'^r$, where $[s^r]$ denotes the operation of extracting the coefficient of $s^r$. The hierarchy introduced in Ref.~\cite{Dong:2026lsb} then takes the form
\begin{equation}
 K_r=\begin{cases}
 \operatorname{Re}\mathcal K_r,&r\text{ odd},\\
 \operatorname{Im}\mathcal K_r,&r\text{ even}.
 \end{cases}
 \label{real-imag-kernel}
\end{equation}
The odd-order equations read $\mathcal K_r\bm C=\bm A^{(r)}$, while
the even-order equations impose homogeneous constraints $\mathcal K_r\bm C=0$ by reality in the symmetric collinear
limit.

To state the counting conjecture, let $\mathcal V_{\leq k}$ denote the row span of
$K_1,\ldots,K_k$. In Ref.~\cite{Dong:2026lsb}, we conjectured that the
number of new independent equation directions appearing at level $k$ is
given by an unsigned Stirling number of the first kind:
\begin{equation}
 \dim\mathcal V_{\leq k}-\dim\mathcal V_{\leq k-1}
 =N_k:=\Stir mk,\qquad 1\leq k\leq m,
 \label{conjecture}
\end{equation}
where $\mathcal V_{\leq0}=0$ and $\Stir mk$ counts permutations of $m$
objects with $k$ cycles. Since $\sum_{k=1}^m N_k=m!$, the retained blocks
are conjectured to provide $m!$ independent equations in total. The even
blocks are homogeneous relations, while the odd blocks involve one-graviton
amplitudes at relative EYM correction order $k-1$; each parity contributes
$m!/2$ equations since $\sum_{k,\rm odd}^mN_k=\sum_{k,\rm even}^mN_k=m!/2$.

For example, at seven points the successive block sizes are
$(6,11,6,1)$, giving cumulative ranks $(6,17,23,24)$. Thus the six
first-order amplitude combinations alone do not determine the 24 YM
entries: they are supplemented by eleven quadratic constraints, six
$F^4$ corrections, and one quartic constraint. At eight points, the
corresponding block sizes are $(24,50,35,10,1)$. 
%We prove the Stirling increments and completeness for an explicit PBW-selected system and construct its explicit inverse. This selected system suffices to reconstruct all $C_\rho$. Identifying its cumulative row spaces with \emph{all} the raw spaces $\mathcal V_{\leq k}$ in Eq.~\eqref{conjecture} requires a further inheritance statement about lower components of higher orders; that stronger claim remains open at general multiplicity.

\textbf{The Poincar\'e--Birkhoff--Witt theorem and period-matrix decoupling.---}We define a recursive map on gluon orderings and identify its coefficients with the monodromy factors in Eq.~\eqref{mixed-disk}. Its commutator structure then allows us to apply the PBW theorem.

%The strict prescription is separate from the PBW projection used below. If a disk coefficient is $c_0+\epsilon c_1+\cdots$ and an adjacent YM amplitude is $C^{(-1)}/\epsilon+C^{(0)}+\cdots$, the full finite disk expression also contains $c_1C^{(-1)}$. The amplitude in the strict prescription retains $c_0C^{(0)}$, with this known pole-mixing contribution separated. Thus $A^{(r)}$ denotes the physical one-graviton amplitude data in the specified strict prescription, before any PBW projection.

To make the algebraic structure explicit, associate a noncommuting letter $x_i$ with each label $i\in Y$. The free associative algebra $T=\mathbb Q\langle x_i:i\in Y\rangle$ consists of linear combinations of words, with multiplication defined to be concatenation . Its multilinear subspace $T_Y$ contains words using each label exactly once; these $m!$ words identify $T_Y$ with the ordering space. For instance, an amplitude vector is represented by $\sum_{\sigma\in S_Y}A^{(r)}(1,\sigma;h)x_{\sigma_1}\cdots x_{\sigma_m}$, after extending scalars to include its kinematic coefficients. Concatenation is an operation on the labels, not multiplication of amplitudes.

To identify the free Lie algebra underlying the monodromy factors, we first express their action on these words recursively. For a nonempty label set $I$, we define
$d_i(I)=s_{iP}+\sum_{j\in I\setminus\{i\}}s_{ij}$. Given the fixed parameter $u=i\pi\alpha'$, we suppress the explicit $u$-dependence of the word-valued maps $D_I$ and define
\begin{equation}
 \begin{aligned}
 D_{\{i\}}(i)&=(e^{u s_{iP}}-1)x_i,\\
 D_I(iv)&=x_iD_{I\setminus\{i\}}(v)-e^{u d_i(I)}D_{I\setminus\{i\}}(v)x_i.
 \end{aligned}
 \label{monodromy-recursion}
\end{equation}
Here $v$ orders the remaining labels in $I\setminus\{i\}$. The recursion places each successive input letter on the left or right of the polynomial constructed from its tail.

To identify this map with the monodromy kernel, consider the coefficient of the reference output word $x_2\cdots x_{n-2}$ for an input $\rho$ ending in $l$. The recursion can produce this output only when the labels below $l$ are placed on the left and those above $l$ on the right. Their input order must therefore be a shuffle of $(2,\ldots,l-1)$ and $(n-2,\ldots,l+1)$ followed by $l$, which is precisely $\rho\in\Omega_l$. There are $n-2-l$ right placements. Their weights multiply to $(-1)^{n-2-l}e^{u[\Phi_l(\rho)-s_{lP}/2]}$, by the definition of $\Phi_l$ in Eq.~\eqref{disk-phase}. Including the seed gives
\begin{equation}
 \begin{aligned}
 &\frac{[x_2\cdots x_{n-2}]D_Y(\rho)}{2i}\\
 &\quad=(-1)^{n-2-l}e^{i\pi\alpha'\Phi_l(\rho)}
 \sin\!\left(\frac{\pi\alpha'}2s_{lP}\right).
 \end{aligned}
 \label{monodromy-identification}
\end{equation}
Here $[x_\sigma]$ extracts the coefficient of $x_{\sigma_1}\cdots x_{\sigma_m}$. Therefore we reproduce both the coefficient and the shuffle restriction in Eq.~\eqref{mixed-disk}, where we have used $e^{u s_{iP}/2}\sin(\pi\alpha's_{iP}/2)=(e^{u s_{iP}}-1)/(2i)$. The relabeling of~\eqref{monodromy-identification} proves the identification for every row, so the matrix of $D_Y$ is $2J(u)$.

The Lie structure is now explicit: for any word $w$, the recursive insertion satisfies
\begin{equation}
 x_iw-e^{u d_i(I)}wx_i
 =[x_i,w]-(e^{u d_i(I)}-1)wx_i.
 \label{monodromy-commutator}
\end{equation}
Since the seed starts at order $u$ and the second term costs an additional power of $u$, the leading order of the monodromy factor is a nested commutator.

The letters and their nested commutators form the free Lie algebra $\mathfrak g\subset T$, with bracket $[f,g]=fg-gf$. The monodromy recursion therefore motivates organizing its output by products of Lie elements. The algebra $T$ is the universal enveloping algebra of $\mathfrak g$. The PBW theorem gives a vector-space isomorphism between $T$ and the symmetric algebra of $\mathfrak g$, realized by symmetrization~\cite{Reutenauer:1986PBW,Reutenauer:1993FreeLie,Brauner:2022Eulerian}. For Lie elements $f_1,\ldots,f_r$, the symmetrization map is 
\begin{equation}
 \Sym(f_1,\ldots,f_r)=\frac1{r!}
 \sum_{\pi}f_{\pi(1)}\cdots f_{\pi(r)}.
 \label{PBWsym}
\end{equation}
Thus every word has a unique decomposition by its number of symmetrized Lie factors. On the multilinear subspace this gives
\begin{equation}
 T_Y=\bigoplus_{r=1}^m V_r,\qquad
 \dim V_r=N_r=\Stir mr,\qquad\sum_{r=1}^mN_r=m!.
 \label{PBWdecomp}
\end{equation}
Here $V_r$ is spanned by symmetrized products of $r$ Lie factors whose disjoint label sets partition $Y$. A block of $k$ distinct labels has $(k-1)!$ independent Lie words; summing their products over partitions into $r$ blocks counts permutations with $r$ cycles, which by definition is the unsigned Stirling number of the first kind.

We define the PBW-degree-$r$ component as the component containing exactly $r$ Lie factors in the PBW decomposition. For example,
\begin{equation}
 x_2x_3=\tfrac12[x_2,x_3]+\Sym(x_2,x_3)
 \label{PBWexample}
\end{equation}
separates a two-letter word into degrees one and two. The PBW symmetrization map is a vector-space isomorphism, not an algebra isomorphism, so concatenation need not preserve PBW degree. We emphasize that PBW degree is not the power $r$ of $\alpha'$ in $A^{(r)}$; their relation is given by the bound below. Let $P_r$ keep the component in $V_r$ and annihilate the others, and let $R_r$ be a rational basis of its independent rows. The explicit forms of $P_r$ and $R_r$ are given in the Supplemental Material (SM); their explicit matrix entries are not needed for the reconstruction formula later.

Returning to Eq.~\eqref{monodromy-commutator}, commutation with $x_i$ does not increase the number of Lie factors, whereas the second term can increase it by at most one and carries at least one additional power of $u$. The seed contains one Lie factor and starts at order $u$. Consequently, the monodromy kernel directly relates the power of $u$, and hence of $\alpha'$, to the PBW degree. Writing $J(u)=\sum_{q\geq1}u^qJ_q$ with $J_q=[u^q]J(u)$, we obtain
\begin{equation}
 \operatorname{im}J_q\subseteq\bigoplus_{p\leq q}V_p,
 \qquad P_rJ_q=0\quad(q<r).
 \label{bound}
\end{equation}
Using the regularity of the period matrix, write $F(u)=\mathds{1}+\sum_{j\geq1}u^jF_j$. Right multiplication by $F_j$ cannot enlarge the image of $J_q$. At order $u^r$, every contribution involving a period
matrix coefficient is of the form $J_qF_{r-q}$ with $q<r$, and therefore has PBW degree strictly below $r$. It is consequently
annihilated by $P_r$, giving
\begin{equation}
 P_r[u^r](JF)=P_rJ_r.
 \label{selection}
\end{equation}
For example, $P_3[u^3](JF)=P_3(J_3+J_2F_1+J_1F_2)=P_3 J_3$ where $P_3$ kills the two terms containing nontrivial period-matrix coefficients. Thus the period matrix drops out after projection onto
PBW degree $r$.

We now define $\widehat{\bm A}_r:=P_r\bm A^{(r)}$. 
In fact, $\bm A^{(r)}$ and $\bm {\widehat{A}}^{(r)}$ coincide for a pure sector, such as the ordinary EYM and the first string correction, but a higher correction can contain lower odd degrees that are removed by $P_r$. We will use shuffle sums of the physical amplitudes to extract the components needed for the solution to avoid an explicit construction of the projector.

Define $W^{(r)}=R_rJ_r$. Its entries are polynomials in the Mandelstam invariants with real coefficients. By construction, $W^{(r)}$ is the coefficient of $u^r$
in $R_rJ(u)$. Restoring $u=i\pi\alpha'$ gives $R_r\mathcal K_r=i^{r-1}\pi^rW^{(r)}$. Hence, for real kinematics, the selected coefficient is real for odd
$r$ and purely imaginary for even $r$. Taking the corresponding real and imaginary parts in Eq.~\eqref{real-imag-kernel} yields
\begin{equation}
 W^{(r)}\bm C=\begin{cases}
 \displaystyle\frac{(-1)^{(r-1)/2}}{\pi^r}R_r\bm A^{(r)},
 &r\text{ odd},\\[2mm]
 0,&r\text{ even}.
 \end{cases}
 \label{selected}
\end{equation}
The reconstruction statement can now be formulated precisely. The square matrix obtained by stacking $W^{(1)},\ldots,W^{(m)}$ is invertible if all $\ell_I\neq0$, and
the first $k$ blocks have rank $\sum_{r\leq k}N_r$. Assuming Eq.~\eqref{input}, the inverse of this stacked system uniquely
reconstructs the subleading YM vector from the odd-order one-graviton amplitudes,
subject to the homogeneous even-order constraints.

\textbf{The explicit solution.---}We now present the reconstruction of the subleading collinear YM amplitudes. To avoid explicitly computing the projector $P_r$, the crucial ingredient is the shuffle sums of one-graviton amplitudes, which eliminate the lower PBW-degree components. Let $\Lambda=\{B_1,\ldots,B_r\}\in\mathfrak F_r(Y)$: take each set partition of $Y$ into $r$ nonempty blocks once, then independently include every internal block ordering with the largest label fixed to the end. A block of $k$ labels contributes $(k-1)!$ orderings; for example, $\{\{2,3,4\},\{5\}\}$ contributes both $234|5$ and $324|5$. Define the $\alpha'^r$--objects:
\begin{equation}
 \mathcal A_r(\Lambda)=\sum_{\sigma\in\Sh(B_1,\ldots,B_r)}
 A^{(r)}(1,\sigma;h).
 \label{sources}
\end{equation}
A sum over $r$ nonempty blocks annihilates every PBW component of degree below $r$. Together with the upper bound in Eq.~\eqref{bound}, this means that Eq.~\eqref{sources} extracts precisely the degree-$r$ information needed for reconstruction. The details of the cancellation are proved in Sec.~\ref{proof:shuffle} of the SM.

The inverse of Eq.~\eqref{selected} can now be written directly in terms of the shuffle sum of the physical amplitudes:
\begin{equation}
 \boxed{\begin{aligned}
 C_\rho={}&\frac{2(-1)^{m+1}}{Q_\rho}
 \sum_{\substack{1\leq r\leq m\\r\ {\rm odd}}}
 \frac{(-1)^{(r-1)/2}}{\pi^r}\\[-1mm]
 &\times\sum_{\Lambda\in\mathfrak F_r(Y)}
 \left[\prod_{B\in\Lambda}\kappa_\rho(B)\right]\cA_r(\Lambda).
 \end{aligned}}
 \label{inverse}
\end{equation}
The reconstruction coefficients in Eq.~\eqref{inverse} have a factorized structure: $Q_\rho$ collects the kinematic poles associated with bipartitions of the gluons, while $\kappa_\rho(B)$ provides a polynomial kinematic factor for each ordered block $B$. We now define these quantities.

For a subset $I$, we define $p_I=\sum_{i\in I}p_i$ and $s_I=p_I^2$ and introduce the bipartition factor
\begin{equation}
 \ell_I=s_I+s_{G\setminus I},\qquad\varnothing\ne I\subseteq Y.
 \label{ell}
\end{equation}
In particular, $\ell_{{i}}=s_{iP}$ and $\ell_Y=s_{1P}$. For a target YM ordering $\rho=\rho_1\cdots\rho_m$, the denominator $Q_\rho$ is constructed from the nested suffix sets
\begin{equation}
 I_j=\{\rho_j,\ldots,\rho_m\},
 \qquad
 Q_\rho=\prod_{j=1}^m\ell_{I_j}.
 \label{Q}
\end{equation}
The block factors $\kappa_\rho(B)$ are constructed from symmetric pair weights defined for $a=\rho_i$ and $b=\rho_j$ with $i<j$ by
\begin{equation}
 t^\rho_{ab}=t^\rho_{ba}
=s_{ab}+\delta_{j,i+1}\ell_{I_{i+1}}.
 \label{weights}
\end{equation}
The second term contributes only when $a$ and $b$ are adjacent in $\rho$. For an ordered block $B=b_1\cdots b_k$, we then define
\begin{equation}
 \kappa_\rho(B)=\prod_{j=1}^{k-1}
 \Bigg(
 \sum_{\substack{l>j\\ b_l\prec_\rho b_j}}
 t^\rho_{b_jb_l}
 -
 \sum_{\substack{l>j\\ b_j\prec_\rho b_l}}
 t^\rho_{b_jb_l}
 \Bigg).
 \label{kappa}
\end{equation}
Here each factor is a signed sum of pair weights, with signs determined by the relative ordering of the labels in $\rho$. In particular, $\kappa_\rho(b)=1$ for a singleton, while $\kappa_\rho(ab)=-t^\rho_{ab}$ for $a\prec_\rho b$.

With these definitions, Eq.~\eqref{inverse} follows by inverting the monodromy recursion and extracting the inverse of the selected PBW system. The recursive cyclic inversions generate the denominator $Q_\rho$, while the nested commutators produce the block factors
$\kappa_\rho(B)$. Note that the additional poles $Q_\rho$ are precisely those predicted in~\cite{Dong:2026kys}. Finally, inserting the physical shuffle sums with the normalization of Eq.~\eqref{selected} yields the reconstruction formula. A proof is given in Sec.~\ref{proof:inverse} of the SM.

\textbf{Examples.---}We now illustrate the reconstruction with explicit examples. At five points, the two YM orderings are determined by a single EYM amplitude. Using $s_{12}+s_{13}+s_{23}=0$, Eq.~\eqref{inverse} gives
\begin{equation}
 \begin{aligned}
 C_{23}&=-\frac{2s_{13}}{\pi s_{12}s_{23}}A^{(1)}(1,2,3;h),\\
 C_{32}&=\frac{2s_{12}}{\pi s_{13}s_{23}}A^{(1)}(1,2,3;h).
 \end{aligned}
 \label{fiveinv}
\end{equation}
These expressions satisfy both the second-order homogeneous relation
$s_{12}^2C_{23}+s_{13}^2C_{32}=0$ and the first-order relation $s_{12}C_{23}-s_{13}C_{32}=2A^{(1)}(1,2,3;h)/\pi$. 

At six points, the EYM amplitudes alone no longer suffice: the reconstruction also requires the first string correction at order $(\alpha')^3$. For brevity, we denote the EYM amplitudes by $E_B=A^{(1)}(1,B;h)$. The $F^4$ correction enters through the sum over all six gluon orderings,
\begin{equation}
 \mathcal A_3(2|3|4)=\sum_{\sigma\in S_{\{2,3,4\}}}A^{(3)}(1,\sigma;h).
 \label{sixsource}
\end{equation}
Consider the YM ordering $234$, momentum conservation reduces the inverse to
\begin{equation}
 \begin{aligned}
 C_{234}=\frac{2}{\pi Q_{234}}\biggl[&
 (s_{13}+s_{14})(s_{14}+s_{24})E_{234}\\
 &+s_{13}s_{24}E_{324}
 -\frac{\mathcal A_3(2|3|4)}{\pi^2}\biggr],
 \end{aligned}
 \label{sixinv}
\end{equation}
where $Q_{234}=s_{234}(s_{12}+s_{34})s_{123}$. Note that for the $F^4$ correction, its six orderings coincide and the sum is six times a single amplitude. This equality can be viewed as a consequence of the Kleiss--Kuijf~\cite{Kleiss:1988ne} (KK)-like relations satisfied by the one-graviton amplitudes, which are discussed in Sec.~\ref{proof:KK} of the SM. Interestingly, in the equivalent expression given in~\cite{Dong:2026lsb}, the same factor of six arises from the $\zeta_2^{-1}$ originated from expanding the string period matrix. 

At seven points, the same mechanism uses sums that preserve the relative order of one pair of gluons. For $i<j$ in $Y=\{2,3,4,5\}$, let $H_{ij}=\mathcal A_3(ij|k|l)$, where $k,l$ are the remaining labels. Each sum contains the twelve orderings in which $i$ precedes $j$. With $t_{ij}=t^{2345}_{ij}$ as defined in Eq.~\eqref{weights}, the YM ordering $2345$ is reconstructed as
\begin{align}
 C_{2345}=-\frac{2}{\pi Q_{2345}}\left[
 \sum_{B\text{ ending in }5}\kappa_{2345}(B)E_B\right.\notag\\
 \left.\hspace{24mm}+\frac1{\pi^2}\sum_{i<j}t_{ij}H_{ij}\right].
 \label{seveninverse}
\end{align}
The first sum contains six EYM amplitudes, with $B$ running over permutations of $2345$ ending in $5$; the second contains six combinations of the $(\alpha')^3$ correction. All coefficients follow directly from Eqs.~\eqref{weights} and \eqref{kappa}. For example,
\begin{align}
 \kappa_{2345}(2345)&=-(t_{23}+t_{24}+t_{25})(t_{34}+t_{35})t_{45},\notag\\
 Q_{2345}&=s_{2345}(s_{12}+s_{345})(s_{123}+s_{45})s_{1234}.
 \label{sevenfactors}
\end{align}

At eight points, amplitudes at order $(\alpha')^5$ enter for the first time, through
\begin{equation}
 \mathcal A_5(2|3|4|5|6)=\sum_{\sigma\in S_Y}A^{(5)}(1,\sigma;h).
 \label{eightsource}
\end{equation}
Note that an individual fifth-order amplitude may contain PBW degrees one, three, and five. Summing its 120 gluon orderings cancels the first two components and extracts the degree-five contribution needed in Eq.~\eqref{inverse}.

\textbf{Twisted-intersection representation.---}
The coefficients in Eq.~\eqref{inverse} admit a realization as
intersections of explicit logarithmic cocycles on
$X={\cal M}_{0,n}\simeq\operatorname{Conf}_m(\mathbb C\setminus\{0,1\})$.
Fix $z_{n-1}=0$, $z_1=1$, $z_n=\infty$, and put
$D_m=dz_2\wedge\cdots\wedge dz_{n-2}$. Introduce the twist $\omega=dW$ through the
multivalued logarithmic potential:
\begin{equation}
 W(z)=
 \sum_{i\in Y}
 \left[\frac{s_{iP}}2\ln z_i+v_i\ln(z_i-1)\right]
 +\sum_{\substack{i<j\\i,j\in Y}}
 s_{ij}\ln(z_i-z_j).
\label{ti:twist}
\end{equation}
The physical Koba--Nielsen twist has  $v_i=s_{1i}$, since
$z_1=1$, while $s_{i,n-1}=s_{iP}/2$ fixes the weights at
$z_{n-1}=0$ \footnote{We temporarily keep the $v_i$ independent to
define the intersection pairing away from the collinear
resonance. Indeed, at the boundary where all moving punctures
approach $z_1$, the twist has residue
\[
 \sum_{i\in Y}v_i+
 \sum_{\substack{i<j\\i,j\in Y}}s_{ij}.
\]
At physical kinematics this equals
$s_G=(\sum_{i\in G}p_i)^2=P^2=0$.
We therefore compute at generic weights and then continue
meromorphically to $v_i=s_{1i}$.}. 
Use the following two Parke--Taylor bases for opposite twists:
\begin{equation}
 \begin{aligned}
 \eta_\rho^+
 &=-\frac{D_m}{
 z_{\rho_m}\prod_{j=1}^{m-1}
 (z_{\rho_{j+1}}-z_{\rho_j})},\\
 \eta_\rho^-&=-\frac{\eta_\rho^+}{z_{\rho_1}-1}.
 \end{aligned}
 \label{ti:PT}
\end{equation}
Write $\theta_\sigma(i,j)=1$ if $i\prec_\sigma j$ and
$\theta_\sigma(i,j)=0$ otherwise, for distinct labels.
The normalized Parke--Taylor intersection
pairing~\cite{Mizera:2017rqa,Mizera:2017cqs,Mizera:2019gea} is the inverse KLT kernel:
\begin{equation}
 \begin{aligned}
 \langle\eta_\rho^-,\eta_\sigma^+\rangle_\omega
 &= ( S^{-1})_{\rho\sigma},\\
  S_{\sigma\rho}
 &=\prod_{i\in Y}\Big\{
 \frac{s_{iP}}2+
 \sum_{j\in Y\setminus\{i\}}
 \theta_\rho(i,j)\theta_\sigma(i,j)s_{ij}
 \Big\}.
 \end{aligned}
 \label{ti:kernel}
\end{equation}
This mixed pairing is independent of $v_i$, permitting the
prescribed collinear continuation.

The cocycles are constructed geometrically from the same
ordered blocks that label the shuffle sources in
Eq.~\eqref{sources}. Recall that a block
$B=b_1\cdots b_k$ represents the right-nested Lie word
$L(B)=[x_{b_1},[\cdots,[x_{b_{k-1}},x_{b_k}]]\cdots]$,
and that $\Lambda=\{B_1,\ldots,B_r\}$ labels the PBW
element $\Sym(L(B_1),\ldots,L(B_r))$.
We first construct the block forms by logarithmic
insertion. Expanding this recursion produces weighted
trees; their Parke--Taylor expansion will then give
the coefficients needed for the intersection pairing.

Using the two collinear punctures $n-1$ and $n$,
define the projectively invariant cross-ratios
and one-forms
\[
 \chi_{ij}=
 \frac{(z_i-z_{n-1})(z_j-z_n)}
      {(z_i-z_n)(z_j-z_{n-1})}=\frac{z_i}{z_j},\ 
 \alpha_i=d\ln\chi_{i1}=d\ln z_i ,
\]
with the second equalities following from our gauge choice, respectively.
Retaining only the four marked points $n-1,n,i,j$
defines a forgetful map
${\cal M}_{0,n}\to {\cal M}_{0,\{n-1,n,i,j\}}\cong {\cal M}_{0,4}$.
The target parametrizes four marked points on the
sphere modulo projective transformations and has
cross-ratio coordinate $\chi_{ij}$. We seek a logarithmic four-point differential with
unit residue at the collision $z_i=z_j$ and symmetric
under exchange of the two collinear punctures.
Requiring no other poles than the three boundary points  of $\overline {\cal M}_{0,4}$ fixes it uniquely to
\begin{equation}
 \beta_{ij}
 =\frac12\,d\ln\frac{(1-\chi_{ij})^2}{\chi_{ij}}\ .
 \label{ti:beta}
\end{equation}
On the compactified four-point moduli curve, this
differential is uniquely determined by having only
simple poles at $\chi=0,1,\infty$, residue $+1$ at
$\chi=1$, and invariance under $\chi\mapsto1/\chi$.
The latter symmetry  exchanges the two collinear punctures $n-1$ and $n$ and fixes
the other two residues to $-1/2$ each. Uniqueness
follows because a difference of two such differentials
would be a holomorphic one-form on $\mathbb P^1$.
For an ordered block $B=b_1\cdots b_k$, construct
its form recursively from the tail
$B'=b_2\cdots b_k$:
\begin{equation}
 \phi_B=
 \begin{cases}
 \alpha_{b_1},&k=1,\\[1mm]
 \displaystyle
 \big(\sum_{a=2}^{k}s_{b_1b_a}\beta_{b_1b_a}\big)
 \wedge\phi_{B'},&k\geq2.
 \end{cases}
 \label{ti:block-recursion}
\end{equation}
On the fiber where the first point $b_1$ of the block
varies, the insertion differential has residue
$s_{b_1b_a}$ at $z_{b_1}=z_{b_a}$, for $a=2,\ldots,k$,
and equal compensating residues
$-\frac12\sum_{a=2}^{k}s_{b_1b_a}$ at $n-1$ and $n$,
with no other poles.
For $\Lambda=\{B_1,\ldots,B_r\}$, let
$\epsilon_\Lambda$ be the sign of the concatenated
block word relative to increasing label order.
Define the forest cocycle by
\begin{equation}
 \Psi_\Lambda^+
 :=(-1)^{m+r+1}2^{1-r}\epsilon_\Lambda\,
   \phi_{B_1}\wedge\cdots\wedge\phi_{B_r}.
 \label{ti:forest}
\end{equation}
Permuting the blocks changes the wedge product and
$\epsilon_\Lambda$ by the same sign, so this
definition is independent of their enumeration.
The overall normalization matches that of the
physical shuffle sources.

To obtain an explicit coordinate expression for the
forest forms in Eq.~\eqref{ti:forest}, we work in the
gauge $(z_{n-1},z_1,z_n)=(0,1,\infty)$ and write
\[
 \lambda_{ij}
 :=s_{ij}\frac{z_i+z_j}{z_i-z_j},
 \qquad
 s_{ij}\beta_{ij}
 =\frac12\lambda_{ij}(\alpha_i-\alpha_j).
\]
Since $\alpha_j\wedge\phi_{B'}=0$ for $j\in B'$,
iterating Eq.~\eqref{ti:block-recursion} gives, for
$B=b_1\cdots b_k$,
\[
 \phi_B=
 2^{1-k}
 \prod_{p=1}^{k-1}
 \left(\sum_{a=p+1}^{k}\lambda_{b_pb_a}\right)
 \alpha_{b_1}\wedge\cdots\wedge\alpha_{b_k}.
\]
Substitution into Eq.~\eqref{ti:forest}, with the
wedge factors reordered into increasing label
order, therefore yields
\begin{equation}
\begin{split}
 \Psi_\Lambda^+
 &=(-1)^{m+r+1}\;\Theta_m
 \prod_{\substack{B=b_1\cdots b_k\\B\in\Lambda}}
 \prod_{p=1}^{k-1}
 \left(\sum_{a=p+1}^{k}\lambda_{b_pb_a}\right),\\
 \Theta_m&=2^{1-m}\bigwedge_{i=2}^{n-2}d\ln z_i.
\end{split} \label{ti:forest-coordinates}
\end{equation}
Empty products are one. Expanding the products amounts to choosing, for each
nonterminal letter $b_p$, one letter $b_a$ to its right
in the same block ($a>p$). Represent the selected
factor $\lambda_{b_pb_a}$ by a directed edge
$b_p\to b_a$. Each complete set of choices produces one tree per
block: following the directed edges always leads
to the last letter, which is the root. Singleton blocks contribute
isolated vertices. Together, these trees form a forest
on $Y$. Thus a fixed block decomposition
$\Lambda=\{B_1,\ldots,B_r\}$ gives a sum of weighted
forests, one for each term in the expanded product.
We apply the partial-fraction identity below to
each complete forest term separately.
For a permutation $\tau$ of $Y$, define
\[
 h_\tau(z)=
 \frac{1}{
 z_{\tau_1}
 \prod_{a=2}^{m}(z_{\tau_a}-z_{\tau_{a-1}})},\ 
 \varepsilon_\tau(i,j)=
 \begin{cases}
 +1,&i\prec_\tau j,\\
 -1,&j\prec_\tau i.
 \end{cases}
\]
For an oriented forest $T$ on $Y$, with
$e=|E(T)|$ edges, the identity reads:
\begin{equation}
 \sum_{\tau\in S_Y}h_\tau(z)\!\!
 \prod_{(i,j)\in E(T)}s_{ij}\;\varepsilon_\tau(i,j)
 =
 \frac{(-1)^e}{\prod_{i=2}^{n-2}z_i}
 \prod_{(i,j)\in E(T)}\!\!\!\lambda_{ij}.
 \label{eq:ti-forest-PF}
\end{equation}
For each directed edge $(i,j)$, define
$\Omega_\sigma(i,j):=s_{ij}\varepsilon_{\sigma^t}(i,j)$,
where $\sigma^t$ is the reversed word $\sigma$.
Using $\eta_\sigma^+=(-1)^m h_{\sigma^t}D_m$
and $e=m-r$, Eq.~\eqref{eq:ti-forest-PF} becomes
\begin{equation}
 \left(\prod_{(i,j)\in E(T)}\lambda_{ij}\right)
 \bigwedge_{i=2}^{n-2}\alpha_i
 =
 (-1)^r\sum_{\sigma\in S_Y}
 \eta_\sigma^+
 \prod_{(i,j)\in E(T)}\!\!\!\Omega_\sigma(i,j),
 \label{eq:ti-tree-PT}
\end{equation}
where $r=m-e$ is the number of connected components.
We now sum over all edge choices within the fixed
blocks of $\Lambda$. For each $\sigma$, these choices
are independent, so the sum factorizes over blocks
and nonterminal letters. The resulting weight of
a block $B=b_1\cdots b_k$ is
\[
 \kappa_\sigma^0(B)
 :=
 \prod_{p=1}^{k-1}
 \left(\sum_{a=p+1}^{k}
 \Omega_\sigma(b_p,b_a)\right),
 \qquad \kappa_\sigma^0(b)=1.
\]
Thus $\kappa_\sigma^0(B)$ denotes the total
ordering weight of a block, rather than an
additional input to the construction.
It is the unshifted version of
Eq.~\eqref{kappa}, obtained by replacing
$t^\sigma_{ij}$ with $s_{ij}$.
Applying this identity to all blocks gives the
Parke--Taylor expansion. The factor $(-1)^r$
in Eq.~\eqref{eq:ti-tree-PT} combines with
$(-1)^{m+r+1}$ in
Eq.~\eqref{ti:forest-coordinates} to give
$(-1)^{m+1}$, so that
\begin{equation}
 \begin{aligned}
 L_{\sigma\Lambda}
 &=2^{1-m}(-1)^{m+1}
   \prod_{B\in\Lambda}\kappa_\sigma^0(B),\\
 \Psi_\Lambda^+
 &=\sum_{\sigma\in S_Y}
       \eta_\sigma^+\ L_{\sigma\Lambda}.
 \end{aligned}
 \label{ti:PT-expansion}
\end{equation}

Equations~\eqref{ti:forest},
\eqref{ti:forest-coordinates}, and
\eqref{ti:PT-expansion} are three representations
of the same rational differential forms, not merely
of the same twisted-cohomology classes. In particular, the geometrically defined
forest forms are logarithmic.
By linearity, Eq.~\eqref{ti:kernel} then gives
\[
 \langle\eta_\rho^-,\Psi_\Lambda^+\rangle_\omega
 =\sum_{\sigma\in S_Y}
   (S^{-1})_{\rho\sigma}L_{\sigma\Lambda}.
\]
The weighted-forest identity established in
Sec.~\ref{proof:forest-intersections} of the SM,
\[
 2^{-m}\sum_{\sigma\in S_Y}
 (S^{-1})_{\rho\sigma}
 \prod_{B\in\Lambda}\kappa_\sigma^0(B)
 =
 \frac1{Q_\rho}
 \prod_{B\in\Lambda}\kappa_\rho(B)
\]
evaluates this matrix product and identifies it with
the reconstruction coefficients in Eq.~\eqref{inverse}:
\begin{equation}
 \langle\eta_\rho^-,\Psi_\Lambda^+\rangle_\omega
 =\frac{2(-1)^{m+1}}{Q_\rho}
   \prod_{B\in\Lambda}\kappa_\rho(B).
 \label{ti:coefficient}
\end{equation}
The geometric construction uses only the unshifted
pair invariants $s_{ij}$; the shifted weights and
channel denominators arise upon evaluating the
intersection pairing.
The reconstruction (\ref{inverse}) therefore becomes \footnote{The construction defines forms for all $m!$
forest labels. In the physical reconstruction,
however, the even-degree equations are homogeneous,
so only the $m!/2$ forms with an odd number of
blocks multiply nonzero source components.
The full family is used in the matrix
factorization proved in the SM.}:
\begin{equation}
 \boxed{
 C_\rho=
 \sum_{\substack{1\leq r\leq m\\r\ {\rm odd}}}
 \frac{(-1)^{(r-1)/2}}{\pi^r}\!\!
 \sum_{\Lambda\in\mathfrak F_r(Y)}
 \langle\eta_\rho^-,\Psi_\Lambda^+\rangle_\omega
 \,\cA_r(\Lambda).
 }
 \label{ti:reconstruction}
\end{equation}

For explicit examples, use (\ref{ti:forest-coordinates}).
At five points the only odd-block forest is $23$,
with
\[
 \Psi_{23}^+
 =s_{23}\beta_{23}\wedge\alpha_3
 =\lambda_{23}\Theta_2.
\]
Equation~\eqref{ti:reconstruction} therefore gives:
\[
 C_\rho=\frac1\pi
 \langle\eta_\rho^-,\Psi_{23}^+\rangle_\omega
 A^{(1)}(1,2,3;h).
\]
For $\rho=23,32$, the intersection numbers are
respectively $-2s_{13}/(s_{12}s_{23})$ and
$2s_{12}/(s_{13}s_{23})$, reproducing
Eq.~\eqref{fiveinv}.

At six points, for example,
\[
 \Psi_{234}^+
 =-(s_{23}\beta_{23}+s_{24}\beta_{24})
   \wedge(s_{34}\beta_{34})\wedge\alpha_4.
\]
The two terms attach label $2$ to either $3$ or $4$,
while label $3$ attaches to $4$.
The three odd-block forest forms required for
reconstruction are
\begin{equation}
 \begin{aligned}
 \Psi_{234}^+
 &=-\lambda_{34}(\lambda_{23}+\lambda_{24})\Theta_3,\\
 \Psi_{324}^+
 &=\lambda_{24}(\lambda_{23}-\lambda_{34})\Theta_3,\quad 
 \Psi_{2|3|4}^+=-\Theta_3.
 \end{aligned}
 \label{ti:sixforms}
\end{equation}
In (\ref{ti:reconstruction}) the three odd-block forms give
\begin{equation}
 \begin{aligned}
 C_\rho={}&
 \frac1\pi\sum_{B\in\{234,324\}}
 \langle\eta_\rho^-,\Psi_B^+\rangle_\omega E_B\\
 &-\frac1{\pi^3}
 \langle\eta_\rho^-,\Psi_{2|3|4}^+\rangle_\omega
 \cA_3(2|3|4),
 \end{aligned}
 \label{ti:sixinverse}
\end{equation}
which reproduces Eq.~\eqref{sixinv} for $\rho=234$.
Thus the higher-order gravitational shuffle sources
are paired with explicit forest cocycles constructed
by logarithmic insertion.
The intersection identities are understood by
meromorphic continuation from generic weights,
without assuming nondegeneracy of the full twisted
cohomology at the resonant collinear point.

\textbf{Discussion and outlook.---} 
We have constructed an explicit all-multiplicity
reconstruction of subleading collinear Yang--Mills
amplitudes from one-graviton EYM amplitudes and their
higher-order string corrections.
The essential step is the identification of the
monodromy kernel in Eq.~\eqref{mixed-disk} with the
weighted Lie recursion \eqref{monodromy-recursion}. The PBW theorem
then organizes the ordering space into sectors of
unsigned Stirling dimensions. Shuffle projections
isolate the highest PBW degree at each order,
eliminating the string-period matrix and yielding
a square system with an explicit factorized inverse (\ref{inverse}).

The reconstruction coefficients also admit a
geometric realization as twisted intersections (\ref{ti:coefficient})
between Parke--Taylor forms and logarithmic forest
cocycles on ${\cal M}_{0,n}$. Their recursive
construction from four-point differentials connects
the PBW block structure to worldsheet geometry.
The factorization (\ref{eq:ti-LMS}), i.e. $LM=S$ relates the selected forward
matrix to the momentum kernel and explains why these
intersections reproduce the inverse coefficients. 
The four-point construction defines the forest
cocycles without reference to the string-period
matrix $F$. Their intersections realize the inverse
of the selected monodromy kernel, whose independence
of the nontrivial period contributions follows
from the PBW projection. Thus the geometric
construction realizes the period-independent
reconstruction, rather than providing a separate
mechanism for eliminating the period matrix. It would therefore be interesting to determine whether
the residue prescription defining the forest
cocycles follows directly from the original
mixed-disk integration contour.

\textbf{Acknowledgments} 
This work is supported by the DFG grant 508889767 {\it 
Forschungsgruppe ``Modern foundations of scattering amplitudes''}. 
The authors used Codex (OpenAI, GPT-6.0 and GPT-6.1) for assistance with literature searches, exploratory calculations, coding, exploring and checking arguments, and language editing. The authors checked all results and arguments and take full responsibility for the scientific content.

% REVTeX selects apsrev4-2 and supplies its longbibliography controls.
\bibliography{Ref}
\onecolumngrid
\clearpage
\setcounter{secnumdepth}{2}
\setcounter{equation}{0}
\setcounter{section}{0}
\setcounter{subsection}{0}
\renewcommand{\theequation}{S\arabic{equation}}
\renewcommand{\thesection}{S\arabic{section}}
\renewcommand{\thesubsection}{\thesection.\arabic{subsection}}
\makeatletter
\renewcommand{\p@subsection}{}
\makeatother
\renewcommand{\theHequation}{supp.\arabic{equation}}
\renewcommand{\theHsection}{supp.\arabic{section}}
\begin{center}
\textbf{\large Supplemental Material}\\[6pt]
\textbf{Poincaré-Birkhoff-Witt Decomposition, Twisted Intersections and Explicit Reconstruction}\\[6pt]
Jin Dong and Stephan Stieberger
\end{center}
This supplement gives the explicit PBW projectors and proves the shuffle selection, inverse reconstruction, and KK-like relations used in the Letter. The monodromy recursion and the PBW degree bound were established in the main text and are used here without repeating their derivations. We retain the Letter's notation: $m=n-3$, $Y=\{2,\ldots,n-2\}$, $G=\{1\}\cup Y$, and $\ell_I=s_I+s_{G\setminus I}$. All amplitude relations use the strict symmetric collinear prescription stated there.

\section{Explicit PBW projectors and independent rows}
For a word $B=b_1\cdots b_k$, define the right-nested Lie element $L(B)=[x_{b_1},[\cdots,[x_{b_{k-1}},x_{b_k}]]\cdots]$, with $L(b)=x_b$. The PBW decomposition of the multilinear ordering space is $T_Y=\bigoplus_{r=1}^m V_r$. A basis of $V_r$ is
\begin{equation}
 b_\Lambda=\Sym(L(B_1),\ldots,L(B_r)),\qquad
 \Lambda\in\mathfrak F_r(Y),
 \label{eq:forestbasis}
\end{equation}
where $\Sym$ averages over all $r!$ orders of the Lie factors. As in the Letter, $\mathfrak F_r(Y)$ runs over each set partition into $r$ nonempty blocks once and all internal block orderings with the largest label fixed to the end. A block with $k$ labels has $(k-1)!$ such orderings. Thus
\begin{equation}
 \dim V_r=\sum_{\substack{\text{set partitions of }Y\\\text{into }r\text{ blocks}}}
 \prod_B(|B|-1)!=\Stir mr=N_r.
\end{equation}
The equality follows by treating each block as a cycle of a permutation; fixing its largest label to the end chooses one representative of that cycle.

For explicit projectors, identify words with position permutations of the increasingly ordered set $Y$, and compose permutations from right to left. The Eulerian projectors are
\begin{equation}
 (P_r)_{\sigma\tau}=[v^r]\binom{v-1+m-\operatorname{des}((\sigma^{-1}\tau)^{-1})}{m},
 \quad P_rP_q=\delta_{rq}P_r,\quad\sum_rP_r=\mathds{1}.
 \label{eq:projectors}
\end{equation}
Here, $\operatorname{des}(\sigma)$ denotes the number of descents of a permutation $\sigma$, i.e. the number of positions $i$ such that $\sigma(i)>\sigma(i+1)$, while $[v^r]$ denotes coefficient extraction with respect to $v$, so that $[v^r]f(v)$ is the coefficient of $v^r$ in the polynomial $f(v)$. We also define $R_r$ to be the nonzero rows of the reduced row-echelon form of $P_r$, using lexicographic word order. Since these rows span the same row space as $P_r$,
\begin{equation}
 R_rP_q=\delta_{rq}R_r,\qquad \operatorname{rank}R_r=N_r.
 \label{eq:rows}
\end{equation}

For $n=5$ with two letters $x_2$ and $ x_3$, we have $R_1=(1,-1)$ and $R_2=(1,1)$. The projector $P_r$ extracts a PBW component of an amplitude vector, whereas $R_r$ selects independent rows of the corresponding equations.

\section{Shuffle sums and PBW-degree selection}
\label{proof:shuffle}
We now prove that the shuffle sums in Eq.~\eqref{sources} annihilate lower PBW degrees and give the normalized coefficients of the surviving component. Let us first define the $r$-fold coproduct on the word algebra by
\begin{equation}
 \Delta_r x_i=\sum_{a=1}^r1^{\otimes(a-1)}\otimes x_i\otimes1^{\otimes(r-a)},
 \qquad \Delta_r(fg)=(\Delta_r f)(\Delta_r g).
\end{equation}
Letters in different tensor factors commute. Expanding a word distributes its letters among the copies while keeping their original order within each copy. Consequently, for any word polynomial $f$,
\begin{equation}
 [x_{B_1}\otimes\cdots\otimes x_{B_r}]\Delta_r f
 =\sum_{\sigma\in\Sh(B_1,\ldots,B_r)}[x_\sigma]f.
 \label{eq:shuffle-pairing}
\end{equation}
Both sides select exactly the words that interleave the specified blocks, with coefficient one.

\paragraph{\normalfont\bfseries Primitivity of Lie elements.}
For two letters, the terms with $x_i$ and $x_j$ in different copies cancel in $[\Delta_r x_i,\Delta_r x_j]$, because those copies commute. Repeating this observation through nested brackets proves
\begin{equation}
 \Delta_r L=\sum_{a=1}^r1^{\otimes(a-1)}\otimes L\otimes1^{\otimes(r-a)}
 \label{eq:primitive}
\end{equation}
for every Lie element $L$. This is the statement that a Lie element is primitive. 

\paragraph{\normalfont\bfseries Cancellation of lower PBW degrees.}
By multiplicativity, a product of $p$ Lie elements occupies at most $p$ copies. If $p<r$, at least one copy is empty in every term of $\Delta_r f$. Since each $B_a$ is nonempty, the coefficient in Eq.~\eqref{eq:shuffle-pairing} is zero. This proves the cancellation for every element of $V_p$, including its symmetrized products.

\paragraph{\normalfont\bfseries Normalization of the degree-$r$ component.}
When $p=r$, each copy must receive exactly one Lie factor with the support of its prescribed block. In the fixed-endpoint basis, $x_B$ has coefficient one in $L(B)$ and zero in the other basis elements on the same support. Each of the $r!$ product orders in the symmetrization therefore gives the same matching tensor coefficient, canceling the factor $1/r!$. To express these statements as matrix identities, index word coordinates by $\sigma\in S_Y$ and forest coordinates by $\Lambda=\{B_1,\ldots,B_r\}\in\mathfrak F_r(Y)$. Define the $m!\times N_r$ matrix $U_r$ and the $N_r\times m!$ matrix $D_r$ by
\begin{equation}
 (U_r)_{\sigma\Lambda}=[x_\sigma]b_\Lambda,
 \qquad
 (D_r)_{\Lambda\sigma}=\begin{cases}
 1,&\sigma\in\Sh(B_1,\ldots,B_r),\\
 0,&\text{otherwise}.
 \end{cases}
 \label{eq:shuffle-matrices}
\end{equation}
Thus $U_r$ expands a linear combination of the symmetrized Lie products $b_\Lambda$ in the word basis, while $D_r$ takes the unweighted shuffle sums of an ordering vector. In particular, $\bm{\mathcal A}_r=D_r\bm A^{(r)}$, with entries $\mathcal A_r(\Lambda)$ defined in Eq.~\eqref{sources}. The cancellation and normalization proved above give
\begin{equation}
 D_rU_r=\mathds{1}_{N_r},\qquad D_rU_p=0\quad(p<r),\qquad
 U_rD_r\sum_{p\leq r}P_p=P_r.
 \label{eq:duality}
\end{equation}
The order bound, Eq.~\eqref{bound} of the Letter, then gives
\begin{equation}
 \widehat{\bm A}_r=P_r\bm A^{(r)}=U_r\bm{\mathcal A}_r.
 \label{eq:reconstruction}
\end{equation}
Thus the unaveraged physical shuffle sums provide exactly the coefficients needed by the inverse, even when individual amplitudes contain lower PBW degrees. The order bound is essential: with more than $r$ factors, several Lie elements can occupy one copy, so an unrestricted higher-degree input need not be annihilated.

\section{Proof of the inverse reconstruction formula}
\label{proof:inverse}
For the target ordering $\rho$ in $C_\rho$, we adopt the notation of the Letter for the suffixes $I_j$, the denominator $Q_\rho=\prod_j \ell_{I_j}$, the pair weights $t^\rho_{ab}$, and the block products $\kappa_\rho(B)$. For brevity, define $\kappa_\rho(\Lambda)=\prod_{B\in\Lambda}\kappa_\rho(B)$.

The proof proceeds in four steps. First, we factor the monodromy recursion into a weighted cyclic difference and a smaller tail map, and invert this factorization recursively; the successive cyclic inversions generate the suffix denominators. Second, comparison between different orderings identifies the accumulated exponential weight with a weighted inversion statistic and determines the pair weights. Third, evaluating this weight on nested commutators yields the block numerators. Finally, a degreewise rescaling followed by the $u\to0$ limit yields the inverse of the selected PBW system.

\paragraph{\normalfont\bfseries Step 1: factor and invert the weighted cyclic recursion.}
We first invert the monodromy map over formal Laurent series in $u$, assuming $\ell_I\neq0$ for every nonempty $I\subseteq Y$. Identify each label word with the corresponding monomial in the word algebra. For $|I|>1$, define two linear maps on $T_I$, the space spanned by words using each label of $I$ exactly once:
\begin{equation}
 T_I(x_i x_w)=\e^{u d_i(I)}x_w x_i,
 \qquad
 A_I(u)(x_i x_w)=x_iD_{I\setminus\{i\}}(u)(w).
 \label{eq:tail-map}
\end{equation}
Here $i\in I$, while $w$ orders the labels in $I\setminus\{i\}$; the $u$ dependence of $T_I$ is left implicit. The map $A_I(u)$ keeps the first letter fixed and applies the smaller monodromy map to the tail. Acting with $\mathds{1}-T_I$ then subtracts the term in which that first letter is cyclically moved to the end, with precisely the weight appearing in Eq.~\eqref{monodromy-recursion}. Hence
\[
 D_I(u)=(\mathds{1}-T_I)A_I(u),
 \qquad
 D_I(u)^{-1}=A_I(u)^{-1}(\mathds{1}-T_I)^{-1}.
\]

The cyclic factor can be inverted explicitly using the geometric-series identity
\[
(\mathds{1}-T_I)
\bigl(\mathds{1}+T_I+\cdots+T_I^{|I|-1}\bigr)
=
\mathds{1}-T_I^{|I|}.
\]
Since every letter is moved once in a full cyclic orbit and
\[
\sum_{i\in I} d_i(I)=\ell_I,
\]
we have
\begin{equation}
 T_I^{|I|}=\e^{u\ell_I}\mathds{1},
 \qquad
 (\mathds{1}-T_I)^{-1}
 =
 \frac{\mathds{1}+T_I+\cdots+T_I^{|I|-1}}
      {1-\e^{u\ell_I}}.
 \label{eq:cyclic}
\end{equation}
The remaining inverse is recursive and acts separately on each first-letter sector:
\[
 A_I(u)^{-1}(x_i f)
 =
 x_iD_{I\setminus\{i\}}(u)^{-1}f,
 \qquad
 f\in T_{I\setminus\{i\}}.
\]
Thus, for a fixed target ordering $\rho=\rho_1\cdots\rho_m$, the rightmost factor $(\mathds{1}-T_I)^{-1}$ first rotates $\rho_1$ to the front, after which $A_I^{-1}$ removes this matched first-letter sector and reduces the problem to the inverse on the tail. Repeating the same procedure fixes $\rho_2,\rho_3,\ldots$ successively. The label sets encountered in this recursive descent are therefore precisely the suffixes $I_j$ defined in the Letter, and each level contributes one factor $(1-\e^{u\ell_{I_j}})^{-1}$.

More explicitly, suppose that at a given stage the target begins with $i$, while the current input word is
\[
 \sigma=W i V .
\]
Exactly one term in the finite cyclic sum in Eq.~\eqref{eq:cyclic} brings $i$ to the front. The resulting tail is $VW$, and the accumulated weight at this stage is
\[
 \exp\left(u\sum_{a\in W}d_a(I)\right).
\]
This motivates the recursive definition
\begin{equation}
 \mathcal E(i\rho',W i V)
 =
 \sum_{a\in W}d_a(I)
 +
 \mathcal E(\rho',VW),
 \qquad
 \mathcal E(i,i)=0,
 \label{eq:Erec}
\end{equation}
where the second term is evaluated on the smaller label set $I\setminus\{i\}$.

At the final one-letter stage,
\[
 D_{\{i\}}(u)^{-1}
\]
contributes
\[
 (\e^{u s_{iP}}-1)^{-1}
 =
 -(1-\e^{u\ell_{\{i\}}})^{-1},
\]
which supplies both the last suffix denominator and the overall minus sign. Iterating the cyclic inversion through all suffixes and using $J(u)=D_Y/2$ therefore gives
\begin{equation}
 (J(u)^{-1})_{\rho\sigma}
 =
 -\frac{2\e^{u\mathcal E(\rho,\sigma)}}
 {\displaystyle\prod_{j=1}^m
 \left(1-\e^{u\ell_{I_j}}\right)}.
 \label{eq:Jinverse}
\end{equation}

\paragraph{\normalfont\bfseries Step 2: identify the exponent as a weighted inversion statistic.}
We next show that the recursively defined exponent admits the explicit representation
\begin{equation}
 \mathcal E(\rho,\sigma)
 =
 \sum_{i<j}t^\rho_{\rho_i\rho_j}
 \mathbf1\!\left\{\sigma^{-1}(\rho_i)>\sigma^{-1}(\rho_j)\right\}.
 \label{eq:inversions}
\end{equation}
Here $\sigma^{-1}(a)$ denotes the position of $a$ in the word $\sigma$, and
$\mathbf1\{\cdots\}$ is the indicator function. Thus each pair
$\rho_i\prec_\rho\rho_j$ contributes precisely when its relative order is
reversed in $\sigma$.

A useful consequence of the recursion \eqref{eq:Erec} is the cyclic-rotation identity
\[
 \mathcal E(\rho,bZ)-\mathcal E(\rho,Zb)
 =
 d_b(I)-\mathbf1\{b=\rho_1\}\ell_I,
\]
where $I$ is the current target suffix. For $b\neq\rho_1$, write
$Z=W\rho_1V$. After the first target deletion, both words $bZ$ and
$Zb$ reduce to the same smaller word $VbW$, so the recursive
contributions coincide and their difference is simply $d_b(I)$.
If instead $b=\rho_1$, moving $b$ from the end to the front traverses
the full cyclic orbit. Using
\[
 \sum_{a\in I}d_a(I)=\ell_I
\]
then gives the additional contribution $-\ell_I$, yielding the stated
identity.

It remains to determine the change of $\mathcal E$ under an adjacent
transposition. Let $a\prec_\rho b$ and compare the two words
$WabV$ and $WbaV$. All recursive steps preceding the target letter $a$
are identical for the two words, so the comparison reduces to the
target suffix $I$ beginning with $a$. After the $a$-step, the remaining
words differ only by a cyclic displacement of $b$. Applying the
rotation identity on $I\setminus\{a\}$ therefore gives
\begin{align*}
 \mathcal E(\rho,WbaV)-\mathcal E(\rho,WabV)
 &=
 d_b(I)-d_b(I\setminus\{a\})
 \nonumber\\
 &\quad
 +\mathbf1\!\left\{
 b\text{ immediately follows }a\text{ in }\rho
 \right\}\ell_{I\setminus\{a\}}
 \\
 &=
 s_{ab}
 +\mathbf1\!\left\{
 b\text{ immediately follows }a\text{ in }\rho
 \right\}\ell_{I\setminus\{a\}}
 \\
 &=t^\rho_{ab}.
\end{align*}
Thus an adjacent inversion of the pair $(a,b)$ changes the exponent
by exactly the corresponding pair weight $t^\rho_{ab}$, including the
additional cut term precisely when $a$ and $b$ are adjacent in the
target ordering.

Both sides of Eq.~\eqref{eq:inversions} vanish for $\sigma=\rho$ and
have the same variation under every adjacent transposition. Since
adjacent transpositions generate all permutations, Eq.~\eqref{eq:inversions}
follows.

\paragraph{\normalfont\bfseries Step 3: evaluate the exponent on Lie forests.}
Now we extend
\[
 \Phi_u(w)=\e^{u\mathcal E(\rho,w)}
\]
linearly from words to the word algebra, with the pair weights
$t^\rho_{ab}$ fixed by the full target ordering $\rho$.
Let $f$ and $g$ have disjoint supports $A$ and $B$, respectively, and define
\[
 c(A,B)
 =
 \sum_{\substack{a\in A,\ b\in B\\
                  a\text{ later than }b\text{ in }\rho}}
 t^\rho_{ab}.
\]
The inversions of a concatenated word $fg$ consist of the inversions
internal to $f$, those internal to $g$, and the cross inversions between
$A$ and $B$. The latter depend only on the two supports. Hence
\[
 \Phi_u(fg)
 =
 \e^{u c(A,B)}\Phi_u(f)\Phi_u(g).
\]
and therefore
\begin{equation}
 \Phi_u([f,g])
 =
 \bigl(
 \e^{u c(A,B)}-\e^{u c(B,A)}
 \bigr)
 \Phi_u(f)\Phi_u(g).
 \label{eq:Phi-commutator}
\end{equation}
At small $u$,
\[
 \e^{u c(A,B)}-\e^{u c(B,A)}
 =
 u\bigl[c(A,B)-c(B,A)\bigr]+O(u^2).
\]
Thus each Lie bracket contributes one additional power of $u$, with
leading coefficient given by the corresponding signed sum of pair
weights.

For a right-nested Lie word
\[
 L(B)
 =
 [x_{b_1},[x_{b_2},\ldots,[x_{b_{|B|-1}},x_{b_{|B|}}]\ldots]],
\]
the outermost bracket separates the first letter from the remaining
support. The factor
$c(\{b_j\},\{b_{j+1},\ldots,b_{|B|}\})
-c(\{b_{j+1},\ldots,b_{|B|}\},\{b_j\})$
is precisely the $j$th signed sum appearing in
Eq.~\eqref{kappa}. Iterating Eq.~\eqref{eq:Phi-commutator} through the
nested brackets therefore gives
\begin{equation}
 \Phi_u(L(B))
 =
 u^{|B|-1}\kappa_\rho(B)+O(u^{|B|}).
 \label{eq:block-leading}
\end{equation}

And thus
\begin{equation}
 \Phi_u(b_\Lambda)
 =
 u^{m-r}\kappa_\rho(\Lambda)+O(u^{m-r+1}),
 \qquad
 \kappa_\rho(\Lambda)=\prod_{B\in\Lambda}\kappa_\rho(B).
 \label{eq:forest-leading}
\end{equation}

\paragraph{\normalfont\bfseries Step 4: extract the inverse of the selected PBW system.}
To isolate the selected system, we remove the leading power $u^r$ from each degree-$r$ component. Define the degree-dependent rescaling operator
\[
 \mathcal R(u)=\sum_{r=1}^m u^rP_r,
 \qquad
 \mathcal R(u)^{-1}=\sum_{r=1}^m u^{-r}P_r,
\]
where the second relation follows from $P_rP_s=\delta_{rs}P_r$ and
$\sum_rP_r=\mathds{1}$. The PBW order bound, Eq.~\eqref{bound}, then implies that
\[
 \mathcal M(u)=\mathcal R(u)^{-1}J(u)
 \quad\hbox{is regular at }u=0,\qquad
 \mathcal M(0)=\sum_{r=1}^mP_rJ_r.
\]
Thus $\mathcal M(0)$ is precisely the selected degree-matched system.

For $u\neq0$, its inverse is
\[
 \mathcal M(u)^{-1}=J(u)^{-1}\mathcal R(u).
\]
We now take the $u\to0$ limit. On any PBW basis element
$b_\Lambda\in V_r$, one has
$\mathcal R(u)b_\Lambda=u^r b_\Lambda$, while
Eqs.~\eqref{eq:Jinverse} and \eqref{eq:forest-leading} give
\begin{equation}
 \lim_{u\to0}\bigl(J(u)^{-1}\mathcal R(u)b_\Lambda\bigr)_\rho
   =\frac{2(-1)^{m+1}}{Q_\rho}\kappa_\rho(\Lambda).
 \label{eq:inverse-limit}
\end{equation}
Indeed, the denominator starts at $(-1)^mu^mQ_\rho$, whereas the
rescaling and the Lie-forest numerator contribute $u^r$ and
$u^{m-r}$, respectively, so all powers of $u$ cancel. Hence
$\mathcal M(u)^{-1}$ has a finite limit on a basis of the full word
space. 

The rows $R_r$ span the row space of $P_r$, so Eq.~\eqref{selected},
with its odd inhomogeneous equations and even imaginary-part
constraints, implies
\[
 \mathcal M(0)\bm C=
 \sum_{\substack{1\leq r\leq m\\r\text{ odd}}}
 \frac{(-1)^{(r-1)/2}}{\pi^r}\widehat{\bm A}_r.
\]
Inserting Eq.~\eqref{eq:reconstruction} and applying
Eq.~\eqref{eq:inverse-limit} gives the inverse formula in the Letter.
Only odd-order gravitational components appear on the right because
the even rows are homogeneous monodromy constraints; no full
even-order amplitude has been set to zero.

\section{KK-like relations for one-graviton amplitudes}
\label{proof:KK}
The one-graviton amplitudes satisfy KK-like relations among gluon orderings at each fixed power of $\alpha'$. These linear relations generalize the ordinary KK-relations for EYM amplitudes to their string corrections. We derive them from the action of the shuffle operator on a fixed PBW degree and the physical reflection identity.

\paragraph{\normalfont\bfseries Action of the shuffle sum.}
The relations can be written compactly using one sum over orderings. We first define an operator $\mathcal T$:
\begin{equation}
 (\mathcal T\bm A^{(r)})[x_\sigma]=\sum_{j=1}^{m-1}
 \sum_{\rho\in\Sh(\sigma_1\cdots\sigma_j,\sigma_{j+1}\cdots\sigma_m)}
 A^{(r)}(1,\rho;h)\,,
 \label{shuffleoperator}
\end{equation}
where on the LHS we  extract the coefficient of
$x_\sigma$. The same definition applies to any ordering vector $a$. In the word algebra, this operator is
\[
 \mathcal T=\mu\Delta-2\mathds{1},
\]
where $\Delta:=\Delta_2$ is the two-fold coproduct introduced above
and $\mu$ concatenates the two tensor factors. The coefficient of
$x_\sigma$ in $\mu\Delta a$ receives contributions from all
decompositions of $\sigma$ into two ordered subsequences and from all
input words obtained by shuffling them. The two trivial decompositions,
in which one subsequence is empty, each reproduce $a(\sigma)$ and are
therefore subtracted.

Each Lie factor is primitive under $\Delta$, and hence is assigned
entirely to one of the two tensor factors. For a symmetrized product of
$p$ Lie factors, there are $2^p$ such assignments. After concatenation
and symmetrization, each assignment reproduces the same PBW element.
Therefore
\begin{equation}
 (\mathcal T+2\mathds{1})b_\Lambda
 =
 2^p b_\Lambda,
 \qquad
 b_\Lambda\in V_p,
\end{equation}
or equivalently
\begin{equation}
 \mathcal T
 =
 \sum_{p=1}^m(2^p-2)P_p.
 \label{eq:shuffle-spectrum}
\end{equation}
Thus the PBW degree-$p$ subspace is an eigenspace of $\mathcal T$ with
eigenvalue $2^p-2$, which gives the numerical coefficients appearing
in the KK-like relations.

\paragraph{\normalfont\bfseries Reflection and the allowed degrees.}
The physical amplitudes obey the reflection relation
\begin{equation}
 A^{(r)}(1,\widetilde\sigma;h)=(-1)^{m+1}A^{(r)}(1,\sigma;h),
 \label{reflection}
\end{equation}
where $\widetilde\sigma$ reverses the gluon labels in $\sigma$. 
Reversal of a Lie word of length $k$ produces the sign
$(-1)^{k-1}$. Hence, on a symmetrized product of $p$ Lie factors
containing all $m$ labels, reversal acts with sign
\[
 (-1)^{m-p}.
\]
The physical reflection identity, Eq.~\eqref{reflection}, requires the
overall sign $(-1)^{m+1}$. Therefore only odd PBW degrees can occur.
Combining this parity condition with the order bound in
Eq.~\eqref{bound} gives
\begin{equation}
 \bm A^{(r)}
 =
 \sum_{p\in S_r}P_p\bm A^{(r)},
 \qquad
 S_r=
 \{\,p\text{ odd}:1\leq p\leq\min(r,m)\,\}.
 \label{alloweddegrees}
\end{equation}

For each $p\in S_r$, Eq.~\eqref{eq:shuffle-spectrum} implies
\[
 [\mathcal T-(2^p-2)\mathds{1}]\,P_p\bm A^{(r)}=0.
\]
Since all factors are polynomials in the same operator $\mathcal T$,
they commute, and the full amplitude therefore satisfies
\begin{equation}
 \boxed{\prod_{p\in S_r}\bigl[\mathcal T-(2^p-2)\mathds{1}\bigr]\bm A^{(r)}=0,}
 \label{KK}
\end{equation}
Here $\mathds{1}$ is the identity and the factors act successively on the amplitude vector. This proves the cumulative KK-like relation. Since the
eigenvalues $2^p-2$ are distinct, the kernel of the corresponding
operator polynomial is
\[
 \bigoplus_{p\in S_r}V_p.
\]
Additional relations may arise from the dynamics within this allowed
PBW subspace.

\paragraph{\normalfont\bfseries Relations at the first few orders.}
The first three odd orders give
\begin{align}
 \mathcal T\bm A^{(1)}&=0,\notag\\
 \mathcal T(\mathcal T-6\mathds{1})\bm A^{(3)}&=0,\notag\\
 \mathcal T(\mathcal T-6\mathds{1})(\mathcal T-30\mathds{1})\bm A^{(5)}&=0
 \qquad(m\geq5).
 \label{KKexamples}
\end{align}
The first line is equivalent to the ordinary KK relations for the EYM amplitudes, while the next two constrain its string corrections. All coefficients are independent of the kinematics. The product in Eq.~\eqref{KK} retains every allowed component: at order $(\alpha')^5$, for example, it annihilates degrees one, three, and five without assuming that the amplitude has a single PBW degree. These are relations among gluon orderings at the same order in $\alpha'$, not equations fixing string corrections from EYM amplitudes. Related KK-like relations for pure-open string amplitudes were studied in Ref.~\cite{Mafra:2021wok}.

\paragraph{\normalfont\bfseries Six-point example: EYM and the $F^4$ correction.}
For $n=6$, the collinear pair is replaced by a graviton $h$, leaving four gluons with labels $1,2,3,4$. Thus we have $m=3$ and $Y=\{2,3,4\}$. Reflection has positive sign and pairs the six gluon orderings as
\begin{equation}
 \begin{aligned}
 A^{(r)}(1,2,3,4;h)&=A^{(r)}(1,4,3,2;h),\\
 A^{(r)}(1,2,4,3;h)&=A^{(r)}(1,3,4,2;h),\\
 A^{(r)}(1,3,2,4;h)&=A^{(r)}(1,4,2,3;h).
 \end{aligned}
 \label{eq:kk-six-reflection}
\end{equation}
The two cuts of $234$ give the shuffles $\Sh(2,34)=\{234,324,342\}$ and $\Sh(23,4)=\{234,243,423\}$. Using reflection, their sum becomes
\begin{equation}
 (\mathcal T\bm A^{(r)})[x_{234}]
 =2\bigl[A^{(r)}(1,2,3,4;h)+A^{(r)}(1,2,4,3;h)
          +A^{(r)}(1,3,2,4;h)\bigr].
 \label{eq:kk-six-shuffle}
\end{equation}
Relabeling gives the same right-hand side for every ordering in $S_Y$.

For EYM, only degree one is present, so $\mathcal T\bm A^{(1)}=0$ gives
\begin{equation}
 A^{(1)}(1,2,4,3;h)
 =-A^{(1)}(1,2,3,4;h)-A^{(1)}(1,3,2,4;h).
 \label{eq:kk-six-eym}
\end{equation}
This is the ordinary four-gluon KK relation with the graviton as a spectator to the color ordering. Together with Eq.~\eqref{eq:kk-six-reflection}, it expresses all six orderings in terms of the two amplitudes with gluons $1$ and $4$ fixed at the ends.

For the first string correction considered here, a single $F^4$ insertion carries the symmetrized trace of four gluon generators, while the graviton carries no color. With gluon $1$ distinguished, this color tensor leaves $\Sym(x_2,x_3,x_4)$, which has PBW degree three. Hence $P_3\bm A^{(3)}=\bm A^{(3)}$ and the relation strengthens to
\begin{equation}
 (\mathcal T-6\mathds{1})\bm A^{(3)}=0.
 \label{eq:kk-six-f4-degree}
\end{equation}
Since the shuffle sum in Eq.~\eqref{eq:kk-six-shuffle} is the same for every ordering, this equation sets six times each amplitude equal to the same quantity. Thus
\begin{equation}
 \begin{aligned}
 &A^{(3)}(1,2,3,4;h)=A^{(3)}(1,2,4,3;h)=A^{(3)}(1,3,2,4;h)\\
 =&A^{(3)}(1,3,4,2;h)=A^{(3)}(1,4,2,3;h)=A^{(3)}(1,4,3,2;h),
 \end{aligned}
 \label{eq:kk-six-f4-equality}
\end{equation}
and the shuffle sum entering the inverse is
\begin{equation}
 \mathcal A_3(2|3|4)=6A^{(3)}(1,2,3,4;h).
 \label{eq:kk-six-factor-six}
\end{equation}
The single-$F^4$ color structure is essential for this stronger conclusion: the cumulative relation $\mathcal T(\mathcal T-6\mathds{1})\bm A^{(3)}=0$ also permits a degree-one component and by itself does not imply equality of the six orderings.

\section{Forest cocycles and twisted intersections}
\label{proof:intersections}

We specify the pairing conventions, verify the
coordinate form of the geometric forest cocycles,
and prove the intersection formula
\eqref{ti:coefficient} using the Parke--Taylor
expansion \eqref{ti:PT-expansion} derived in the
main text.

\subsection{Pairing conventions and collinear specialization}
\label{proof:intersection-conventions}

Let $\nabla_\omega=d+\omega\wedge$, where
$\omega=dW$ and $W$ is defined in
Eq.~\eqref{ti:twist}.
We regard the first and second entries of the pairing
as classes in $H^m(X,\nabla_\omega)$ and
$H^m(X,\nabla_{-\omega})$, respectively.
Thus the superscripts on $\eta_\rho^\pm$ distinguish
the two chosen Parke--Taylor families; they do not
specify the sign of the connection.
For logarithmic forms $fD_m$ and $gD_m$, our
normalization is fixed by the critical-point formula
\cite{Mizera:2017cqs}
\begin{equation}
 \langle fD_m,gD_m\rangle_\omega
 =
 (-1)^m\sum_{E(z)=0}
 \frac{f(z)g(z)}
 {\det\bigl(\partial E_i/\partial z_j\bigr)},
 \label{eq:ti-critical}
\end{equation}
where the sum runs over the isolated nondegenerate
solutions of
\begin{equation} 
E_i(z):=\frac{\partial W}{\partial z_i}
 =\frac{s_{iP}}{2z_i}
 +\frac{v_i}{z_i-1}
 +\sum_{\substack{j\in Y\\j\ne i}}
 \frac{s_{ij}}{z_i-z_j}
 =0\ .\label{eq:ti-scattering}
 \end{equation}
At generic weights, these are the critical-point
equations of $W$, or equivalently the scattering
equations associated with the auxiliary weights.
At $v_i=s_{1i}$ they take the physical collinear
form. The pairing at this resonant specialization
is obtained by meromorphic continuation of the
generic result.

With the signs and word conventions of
Eq.~\eqref{ti:PT}, the Parke--Taylor intersection
formula gives
\begin{equation}
 G_{\rho\sigma}
 :=\langle\eta_\rho^-,\eta_\sigma^+\rangle_\omega
 =(S^{-1})_{\rho\sigma}.
 \label{eq:ti-G}
\end{equation}
One way to match conventions is to use the
common-tree expansion of the mixed Parke--Taylor
pairing. It reduces to the double-partial rooted
current with words
$(n-1)\widetilde\rho$ and
$(n-1)\widetilde\sigma$, where a tilde reverses a word.
The inverse of this current matrix is the momentum
kernel
\[
 S_{\sigma\rho}
 =
 \prod_{i\in Y}
 \left[
 \frac{s_{iP}}2+
 \sum_{\substack{j\in Y\\
                 i\prec_\rho j,\ i\prec_\sigma j}}
 s_{ij}
 \right].
\]
The minus sign in $\eta_\sigma^+$ in
Eq.~\eqref{ti:PT} fixes the relative sign of this
mixed pairing. In particular, for a single moving
label it gives $G_{ii}=2/s_{iP}$.

The common-tree propagators involve only labels in
$\{n-1\}\cup Y$, with pair weights
$s_{i,n-1}=s_{iP}/2$ and $s_{ij}$.
Consequently $G$ is independent of the auxiliary
weights $v_i$.
The physical specialization $v_i=s_{1i}$ is
resonant: the residue at the boundary where all
moving points approach $z_1$ is
\[
 \sum_{i\in Y}v_i+
 \sum_{\substack{i<j\\i,j\in Y}}s_{ij}
 =s_G=0.
\]
Equation~\eqref{eq:ti-G} nevertheless defines the
meromorphic continuation of the selected mixed
pairing to this locus, away from its remaining
kinematic poles. No nondegeneracy statement about
the full resonant twisted cohomology is required.

\subsection{Coordinate form of the geometric recursion}
\label{proof:forest-partial-fractions}

We derive Eq.~\eqref{ti:forest-coordinates}
from the geometric definition
\eqref{ti:forest}. In the fixed gauge, we use
the identity:
\[
 s_{ij}\beta_{ij}
 =\frac12\lambda_{ij}(\alpha_i-\alpha_j),
 \qquad \alpha_i=d\log z_i.
\]
For a block $B=b_1\cdots b_k$, the form (\ref{ti:block-recursion}) on its tail
has top degree in the tail coordinates. Hence
$\alpha_j\wedge\phi_{b_2\cdots b_k}=0$ for every
$j\in\{b_2,\ldots,b_k\}$.
Induction in Eq.~\eqref{ti:block-recursion} gives
\begin{equation}
 \phi_B=
 2^{1-k}
 \prod_{p=1}^{k-1}
 \left(\sum_{a=p+1}^{k}\lambda_{b_pb_a}\right)
 \alpha_{b_1}\wedge\cdots\wedge\alpha_{b_k}.
 \label{eq:ti-block-product}
\end{equation}
For a forest with $r$ blocks, these factors contribute
$2^{r-m}$. Multiplication by the prefactor
$2^{1-r}$ in Eq.~\eqref{ti:forest} leaves $2^{1-m}$ entering (\ref{ti:forest-coordinates}).
Reordering the block wedges into increasing label
order contributes $\epsilon_\Lambda$, cancelling
the explicit $\epsilon_\Lambda$ in that definition.
This proves the coordinate forest formula (\ref{ti:forest-coordinates}), including its sign and normalization.

\subsection{The forward system in block coordinates}
\label{proof:forest-forward}

Our aim is to prove $LM=S$, where $L$ contains
the forest coefficients in
Eq.~\eqref{ti:PT-expansion} and $S$ is the momentum
kernel in Eq.~\eqref{ti:kernel}.
This identifies the intersection matrix $S^{-1}L$
with the inverse of the selected forward system.

Using the shuffle extraction matrices of
Eq.~\eqref{eq:shuffle-matrices}, define
\begin{equation}
 M_{\Lambda\rho}:=(D_rJ_r)_{\Lambda\rho},
 \qquad \Lambda\in\mathfrak F_r(Y),
 \quad 1\leq r\leq m.
 \label{eq:ti-M}
\end{equation}
The column label $\rho$ specifies an input ordering,
while the row label $\Lambda$ specifies the blocks
of a shuffle sum.
Collecting all degrees gives a square
$m!\times m!$ matrix.
The selection and normalization proved in
Sec.~\ref{proof:shuffle} identify this matrix with
the forward system inverted in
Sec.~\ref{proof:inverse}.

To compute $LM$, we describe the selected recursion
in terms of blocks of labels.
Fix an ordering $\rho$ and put
\[
 w_\rho(i)
 :=s_{iP}+\sum_{j:\,i\prec_\rho j}s_{ij}.
\]
This is the weight $d_i(I)$ in
Eq.~\eqref{monodromy-recursion} when $I$ consists
of $i$ and all labels to its right in $\rho$.

The recursion processes the labels from right
to left. After the degree selection, inserting
a label $i$ has two possibilities:
it joins one of the existing blocks, or it starts
a new singleton block with weight $-w_\rho(i)$.
These are the two contributions in
Eq.~\eqref{monodromy-commutator} at equal order
in $u$ and PBW degree.
Terms of higher order in the exponential do not
survive this selection.

Starting from the rightmost label, these choices
generate every set partition of $Y$ exactly once.
The order within each block is inherited from
$\rho$. We write
\[
 \pi=\{I_1,\ldots,I_r\}\in\Pi_r(Y),
 \qquad
 B_a=\rho|_{I_a}=b_{a,1}\cdots b_{a,k_a},
\]
where $\Pi_r(Y)$ denotes the partitions into
$r$ nonempty subsets and $\rho|_{I_a}$ retains
only the labels in $I_a$. Here $\pi$ records only the subsets of the partition;
their internal order is supplied by $\rho$.
This differs from the fixed-endpoint block labels
$\Lambda$ indexing the rows of $M$.

The last label $b_{a,k_a}$ of each block is
the first one encountered in that block.
It therefore supplies its root weight
$w_\rho(b_{a,k_a})$.
For the rightmost label $\rho_m$, the seed weight
is $s_{\rho_m P}=w_\rho(\rho_m)$.
Starting the other $r-1$ blocks contributes
$-w_\rho(i)$ each, giving the overall sign
$(-1)^{r-1}$.
Together with $J=D_Y/2$, the forward contribution
of a partition is thus
\[
 \frac{(-1)^{r-1}}2
 \prod_{a=1}^{r}w_\rho(b_{a,k_a}).
\]

The matrix $L$ supplies the corresponding block
weights $\kappa_\sigma^0(B_a)$ from
Eq.~\eqref{ti:PT-expansion}.
They can be evaluated directly in the ordering
inherited from $\rho$, even if a block does not
end in its largest label.%
\footnote{This evaluation is independent of the
fixed-endpoint Lie basis used to define the shuffle
coordinates. For disjoint supports $A,B$, introduce
formal symbols $e_A$ and set
$[e_A,e_B]=\Omega_\sigma(A,B)e_{A\cup B}$, with
$\Omega_\sigma(A,B)=
\sum_{i\in A,j\in B}\Omega_\sigma(i,j)$.
Antisymmetry and additivity of $\Omega_\sigma$
imply the Jacobi identity on disjoint supports.
Under $x_i\mapsto e_{\{i\}}$, the nested Lie
element $L(B)$ evaluates to
$\kappa_\sigma^0(B)e_{\operatorname{supp}B}$.
Hence these block weights respect the relations
among Lie elements and may be applied before
expansion in the fixed-endpoint basis.
The normalization of the resulting block
coefficients is fixed by
Eq.~\eqref{eq:duality}.}
We can therefore compute $LM$ directly by summing
over partitions, with one root weight from $M$
and one block weight from $L$ for each component.

\subsection{Summing the forests}
\label{proof:forest-factorization}

Combining the weights just described gives
\begin{equation}
 (LM)_{\sigma\rho}
 =
 2^{-m}\sum_{r=1}^{m}(-1)^{m+r}
 \sum_{\pi\in\Pi_r(Y)}
 \prod_{a=1}^{r}
 \left[
 w_\rho(b_{a,k_a})\,
 \kappa_\sigma^0(B_a)
 \right].
 \label{eq:ti-block-contraction}
\end{equation}
The prefactor combines $(-1)^{r-1}/2$ from the
forward recursion with
$2^{1-m}(-1)^{m+1}$ from
Eq.~\eqref{ti:PT-expansion}.

We now expand each block weight into trees,
exactly as in the discussion after
Eq.~\eqref{ti:forest-coordinates}.
For every label except the last one in a block,
choose one directed edge to a label to its right
in that block.
The last label is the root.
Since the block ordering is inherited from $\rho$,
all edges also point to the right in $\rho$.

A partition into $r$ blocks gives a forest with
$r$ roots and $m-r$ edges.
Because $(-1)^{m+r}=(-1)^{m-r}$, the sign in
Eq.~\eqref{eq:ti-block-contraction} can be assigned
one minus sign per edge.
Thus the weights of an individual forest are
\[
 \text{root }i:\quad w_\rho(i),
 \qquad
 \text{edge }i\to j:\quad-\Omega_\sigma(i,j),
 \quad i\prec_\rho j.
\]

Instead of first choosing the blocks and then
their trees, we may construct the same forests
directly on $Y$:
each label $i$ either is a root or has one outgoing
edge to a label $j$ to its right in $\rho$.
These choices are independent.
No cycle is possible, and each connected component
has exactly one root.
Conversely, the components of the resulting forest
uniquely recover its block partition.
This counts exactly the terms in
Eq.~\eqref{eq:ti-block-contraction}, with no
additional multiplicity.

The sum over forests is consequently a product
over the choices at each label:
\begin{equation}
 (LM)_{\sigma\rho}
 =
 2^{-m}\prod_{i\in Y}
 \left[
 w_\rho(i)
 -\sum_{j:\,i\prec_\rho j}\Omega_\sigma(i,j)
 \right].
 \label{eq:ti-root-parent}
\end{equation}
Using the definition of $w_\rho$,
\[
 w_\rho(i)
 -\sum_{j:\,i\prec_\rho j}\Omega_\sigma(i,j)
 =
 s_{iP}
 +\sum_{j:\,i\prec_\rho j}
 \bigl(s_{ij}-\Omega_\sigma(i,j)\bigr).
\]
For $i\prec_\sigma j$ the term in parentheses
equals $2s_{ij}$; for $j\prec_\sigma i$ it vanishes.
Therefore
\begin{equation}
 (LM)_{\sigma\rho}
 =
 \prod_{i\in Y}
 \left[
 \frac{s_{iP}}2+
 \sum_{\substack{j\in Y\\
                 i\prec_\rho j,\ i\prec_\sigma j}}
 s_{ij}
 \right]
 =S_{\sigma\rho},
 \qquad LM=S.
 \label{eq:ti-LMS}
\end{equation}
The product is precisely the momentum kernel
of Eq.~\eqref{ti:kernel}.
The identity holds polynomially in the weights,
before imposing the physical kinematic
specialization.
It evaluates the matrix product $S^{-1}L$
appearing after Eq.~\eqref{ti:PT-expansion}
and supplies the link to the intersection formula
\eqref{ti:coefficient}, as detailed below.

\subsection{Reconstruction as a twisted pairing}
\label{proof:forest-intersections}

The Parke--Taylor pairing
$S^{-1}$ in Eq.~\eqref{eq:ti-G} and the cocycle
expansion \eqref{ti:PT-expansion} give, in matrix form,
\begin{equation}
 \bigl(\langle\eta_\rho^-,\Psi_\Lambda^+\rangle_\omega
 \bigr)_{\rho,\Lambda}
 =S^{-1}L=M^{-1}.
 \label{eq:ti-intersection-proved}
\end{equation}
The last equality follows from $LM=S$ at generic
kinematics. Thus the geometrically constructed
cocycles realize the inverse of the selected
shuffle-source system.

The explicit inverse already obtained in
Eq.~\eqref{eq:inverse-limit}, together with the
shuffle normalization of Sec.~\ref{proof:shuffle},
has entries
\[
 (M^{-1})_{\rho\Lambda}
 =\frac{2(-1)^{m+1}}{Q_\rho}
   \prod_{B\in\Lambda}\kappa_\rho(B).
\]
Substitution into
Eq.~\eqref{eq:ti-intersection-proved} proves
Eq.~\eqref{ti:coefficient} in the main text;
inserting the odd-degree sources of
Eq.~\eqref{sources} then gives
Eq.~\eqref{ti:reconstruction}.
Equivalently, using $L$ from
Eq.~\eqref{ti:PT-expansion} yields the
weighted-forest identity used immediately before
Eq.~\eqref{ti:coefficient}:
\[
 2^{-m}\sum_{\sigma\in S_Y}
 (S^{-1})_{\rho\sigma}
 \prod_{B\in\Lambda}\kappa_\sigma^0(B)
 =
 \frac{1}{Q_\rho}
 \prod_{B\in\Lambda}\kappa_\rho(B).
\]
The unshifted weights in $L$ thereby produce the
shifted weights and channel denominators of the
inverse through the pairing with $S^{-1}$.
These rational identities extend to the physical
collinear locus by the prescription of
Sec.~\ref{proof:intersection-conventions}.

The construction determines the cocycles relative
to the distinguished punctures, ordered PBW blocks,
and prescribed residues. It does not establish
that this residue prescription is uniquely selected
by the original mixed-disk integration contour.

\end{document}